\def\({\left(}
\def\[{\left[}
\def\l\{{\left\{}
\def\){\right)}
\def\]{\right]}
\def\r\}{\right\}}
\def\what{\widehat}
\def\raw{\rightarrow}
\def\bA{\bar A}
\def\am{\bar A_{\rm max}}
\def\cf{{\cal F}}

\def\etal{{\sl et al.} }
\def\ratio#1#2{{{#1}\over{#2}}}
\def\KH{Kelvin-Helmholtz }
\def\3d{three-dimensional }
\def\2d{two-dimensional }
\def\be{\begin{equation}}
\def\ee{\end{equation}}
\def\bef{\begin{figure}}
\def\ef{\end{figure}}

\def\DXDYCZ#1#2#3{\left({\partial#1\over\partial#2}\right)_{#3}}
\def\lapp{\mathbin{\raise2pt \hbox{$<$} \hskip-9pt \lower4pt
\hbox{$\sim$}}}
\def\gapp{\mathbin{\raise2pt \hbox{$>$} \hskip-9pt \lower4pt
\hbox{$\sim$}}}

\documentclass{aa}
\usepackage{graphicx}
\begin{document}

   \thesaurus{02.08.1; 02.09.1; 08.05.1; 09.10.1}
   \title{On the matter entrainment by stellar jets  
and the acceleration of molecular outflows}

   \author{ M. Micono\inst{1,2}
   \and G. Bodo\inst{1} \and S. Massaglia\inst{2} \and P. Rossi\inst{1}\and A. Ferrari\inst{2}          }

\institute{
Osservatorio Astronomico di Torino, Viale Osservatorio 20, I-10025
Pino Torinese, Italy \and
Dipartimento di Fisica Generale dell'Universit\`a,
Via Pietro
Giuria 1, I-10125 Torino, Italy
 }

   \offprints{M. Micono}

   \date{Received; accepted }

\maketitle

\begin{abstract}
We study, by numerical simulations, the entrainment process
in a supersonic, radiative jet flow, during the evolution of
Kelvin-Helmholtz instabilities, in the context of the 
the acceleration problem
of molecular bipolar outflows, observed in
Giant Molecular Clouds. 
Our results show that a large fraction of the initial jet momentum 
can be transferred to the ambient medium by this process. 
We therefore analyze in  detail the instability evolution 
and compare some of the
main observational properties of molecular outflows with 
those of the entrained material that we observe in
our simulations. In particular, we find a good agreement for
the mass vs. velocity distribution and for the outflow collimation
structure, especially when a light jet is moving into 
a denser ambient medium. 
This is probably the case for (obscured) optical jets driving 
powerful molecular outflows in the denser environment of
the inner regions of molecular clouds.

\keywords{Hydrodynamics --
                instabilities -- stars: early-type --  ISM: jets
          and outflows }
   \end{abstract}
\section{Introduction}

Young Stellar Objects (YSO) are accompanied by different kinds of 
outflow phenomena, whose most striking manifestations are optical 
jets and molecular outflows; however the mutual relationships 
existing between such different kinds of outflows still remain 
to be clarified.  In the past, the prevailing idea was that fast 
atomic winds, with low collimation, were the primary accelerating 
agents of molecular outflows (Lada 1985; Lizano et\ al.\ 1988), 
while optical jets were considered as parallel  phenomena
with no dynamical significance, because their momentum 
transport rate was estimated to be much lower than that of molecular
outflows. These models, however, failed to explain, in a satisfactory
way, 
some important outflow properties, while, on the opposite, models based 
on highly collimated flows appeared to be able to account for them
  (Masson \& Chernin 1992; Stahler 1994). Additional problems to the 
wind models have been posed by the discovery of molecular outflow
components
with a higher degree of collimation and higher velocities with respect
to the standard component, such as EHV components and molecular bullets
(Bachiller 1996). Moreover,the 
estimates of the momentum carried by optical jets have been
reconsidered: 
more accurate
comparisons of the observed emission with shock models has allowed
to show that jets have a low degree of ionization (Raga, Binette \&
Cant\'o
1990) and therefore 
their density is much higher than what was previously thought and, 
secondly, the time scales of the jet phenomenon appear also to be
much longer than what was previously thought (Parker, Padman \& Scott
1991;
Bally \& Devine 1994); therefore the total momentum that optical jets
can deposit in the ambient medium is much larger than what was
previously
estimated and  the problem of momentum deficit seems to be overcome.
As a consequence of all these considerations, there is a growing
consensus 
that there must exist a causal connection between optical jets and
molecular
outflows (Bachiller 1996), however the  physical mechanism through
which jets drive the outflows have not yet been clarified.    
The proposed models fall 
essentially into two classes: in one case the ambient material is 
entrained in a turbulent mixing layer along the length of the jet
(sometimes this is referred to as steady state entrainment),
while, in the second case, the entrainment is performed by 
a working surface, either the leading bow shock or internal
working surfaces due to variability in the jet emission (sometimes
this is referred to as prompt entrainment).

While there are several studies concerning the second mechanism 
(see Cabrit et al. 1997 and references therein),
for the first one there are only a few attempts to model the 
physical processes involved (Stahler 1994, Taylor \& Raga 1995). 
Both mechanisms should be at work: working surfaces and
bow shocks are surely present in jets and the development of a
turbulent mixing layer between jet and ambient material seems
unavoidable as the result of the evolution of Kelvin--Helmholtz 
instabilities. Although in several situations  the prompt 
entrainment mechanism seems to fit very well
the observational properties (see e.g.\ Davis et al. 1997),
a thorough study of the steady state entrainment is also needed in order
to assess its importance. In any case, more work is needed in the 
theoretical analysis of both 
mechanisms since they both seem to fail in explaining the large size 
of molecular outflows as compared to that of optical jets.
In this paper we address the problem of steady state
entrainment;   
its study is obviously 
complicated by the fact that it involves turbulence and one possible
approach, which we will follow,
 is the use of direct numerical simulations. 
We have studied the evolution of Kelvin--Helmholtz instabilities
in several different conditions both in two and in three dimensions
(Bodo et al. 1994, 1995, 1998, Rossi et al. 1997, 
Micono et al. 1998, 2000)
and in this paper we analyze the results of three--dimensional 
simulations in a radiative situation. The study of a fully
three-dimensional
situation is essential for analyzing the entrainment properties, since
it is well known that the properties of turbulence are very different in
2D and in 3D (for a comparison of 2D and 3D results on the evolution of 
Kelvin-Helmholtz instabilities see e.g. Bodo et al.\  1998).
Following the above 
considerations, the focus of our analysis will
be  on the properties of the entrainment of the ambient material, 
while a more general discussion on the evolution of the instability
is presented in Micono et\ al.\ (2000) (hereinafter referred as 
Paper I).

The plan of the paper is the following: after a summary of the properties
of molecular outflows, given in Sect. 2, we present our model in Sect. 3, and 
our results in Sect. 4. 
Finally, a summary and the conclusions are given in Sect. 5.

\section{Properties of molecular outflows}

Molecular outflows consist of a pair of oppositely-directed
and poorly collimated lobes, symmetrically located about an embedded
YSO, detected in broad mm-wave emission lines, especially of CO.
Their maximum sizes range from 0.04 to 4 pc, the
dynamical ages from $10^3$ to 2 $\times 10^5$ yr,
H$_2$ densities are $\sim 10^5$ cm$^{-3}$,
their total momentum is in the range 0.1 -
1000 M$_\odot$ km s$^{-1}$ (Bachiller 1996, Fukuy et al. 1993). 
In general, they are poorly collimated, with ratios of major 
to minor axis of 3-10, although highly collimated molecular outflows 
have been detected (NGC 2264G, Lada \& Fich 1996, Mon R2, Meyers-Rice 
\& Lada 1991, RNO 43, Padman et al. 1997). 
Average velocities observed in molecular outflows are in the range 
$10-30$km s$^{-1}$, but high velocity flows are also observed, with
bulk velocities up to 100 km s$^{-1}$ (eg. IRc2, Rodriguez-Franco et al. 1999).
Also in low-velocity outflows it is possible to find clumps of
Extremely High Velocity gas (the so-called EHV bullets), which
shows up as a bump in the tails of spectral lines (Cabrit et al. 1997);
 the mass involved in EHV gas is a very small fraction of the total 
flow mass, from $10^{-4}$ to $10^{-2}$ M$_\odot$.

There are a number of properties of the outflows with which theoretical
models for their acceleration must confront. First we have the 
 distribution of  the flow mass 
with velocity, which has typically the form of             
a power law $M(v) \propto v^{\gamma}$, with a break at high velocities.
The values of  $\gamma$ range from $-1.3$ to $-2.1$ (Cabrit et al. 1997,       
Masson \& Chernin 1992). The break is found at velocities of 
$\sim 25-30$ km s$^{-1}$ and the distribution    
at higher velocities is steeper with slope $\sim$ $-3.5$, $-5$.

We have then the collimation properties: 
the flow collimation increases with  velocity, in fact 
the higher velocity material tends to lie in the center of the outflow, while  
limb-brightening is strongest at low-velocities
(L1551-IRS5 Uchida et al.  1987, Moriatry-Schieven \& Snell 1988; 
NGC 2071 Moriatry-Schieven, 
Hughes \& Snell 1989; NGC 2024 Richer et al.\  1992; NGC 2264G Margulis
et al.\  1990). In addition molecular outflows show a high degree of 
bipolarity: in the same lobe
the occurrence of overlapping blue-shifted and red-shifted emission
is rare: the contrast in intensity 
 between red and blue-shifted emission increases with velocity   
up to values $\gapp 20$ for high velocity gas (Lada \& Fich 1996),
and this implies a large predominance of longitudinal  motions      
(Cabrit et al. 1997, Lada \& Fich 1996).

Many outflows show an apparent linear acceleration, i.e. 
the largest velocities in the molecular material are found farthest 
from the star.  Molecular outflows not showing this property are
also observed, for example in NGC 2024 the velocity is 
constant over 75\% of the outflow (Padman et al. 1997).

Finally, Chernin \& Masson (1995) studied the distribution 
of mass $dM/dz$ and momentum $dP/dz$ as a function of distance along
the flow axis and they found that both distributions peak
in the middle of the lobe, with minima near the star and at the 
extremity of the flow.
The same trend is observed for the cross sectional area of the flow 
(i.e. the lobes are widest at the location of the momentum peak).

\section{Numerical Simulations}

We have studied the entrainment properties of a supersonic jet: the turbulent
mixing layer, through which the external matter is entrained by the jet
motion, is formed by the growth and evolution of Kelvin--Helmholtz 
instabilities and, therefore, in our simulations we have given to the
jet an initial small perturbation and we have then followed the flow 
evolution. The details of the physical and numerical setup are given in Paper
I, and here we summarize the main characteristics. Since we study the
evolution over long times, we perform the instability analysis  
 using the so called temporal approach, in which one studies the
temporal evolution of the instability in a section of an infinite periodic 
jet (periodic boundary conditions are used at the longitudinal boundaries).
The initial velocity and density profiles are given by the functional forms
$$
V_x (y,z) = V_{_0} ~ {\rm sech} \left[ \left( \sqrt{y^2 + z^2} \over
a\right)^w\right]
\,,
$$
$$
{ {\rho_{_0} (y, z; \nu)} \over {\rho_{\rm jet}}} = \nu - (\nu - 1) ~ {\rm
sech} \left[ \left( \sqrt{y^2 + z^2} \over a\right)^w\right]
\,,
$$
where $V_{_0}$ is the velocity on the jet axis, $a$ is the initial jet
radius,  $m$ is a parameter controlling the interface (shear) layer
width (typically, we set $w=8$) and $\nu$ is the ratio of the density 
at $r= \sqrt{y^2 + z^2} = \infty$,
$\rho_{\infty}$, to that on the jet axis ($y = z = 0$) at $t = 0$,
$\rho_{\rm jet} \equiv \rho_{_0} (y=0, z=0)$. The perturbations have
the form 
$$
V_y(x,y,z) = {V_{y,_0} \over n_{_0}} ~ {\rm sech} \left( \sqrt{y^2 + z^2} \over
a\right)^w \times
$$
$$
 {\sum_{n=1}^{n_{_0}}} \sin (nk_{_0}x + \phi_n) \,,
\eqno(2a)
$$
$$
V_z(x,y,z) = {V_{y,_0} \over n_{_0}} ~ {\rm sech} \left(  \sqrt{y^2 + z^2} \over
a\right)^w \times
$$
$$
 {\sum_{n=1}^{n_{_0}}} \cos (nk_{_0}x + \phi_n) \,,
\eqno(2b)
$$
where  $V_{y,_0} = 0.005 V_{_0}$ is the amplitude of the initial perturbation
and $\{\phi_n\}$ are the phase shifts of the various Fourier components.
The perturbations are thus given by a superposition of 
a number $n_0 =8$ of longitudinally 
periodic transverse velocity disturbances, in order to excite a wide range of
modes. In order to study the entrainment and mixing properties we also
follow the evolution of a tracer, a scalar field, $\cal T$, passively 
advected by the 
fluid. In the initial configuration $\cal T$ is put equal to one inside the 
jet and to zero outside, this allows to distinguish the jet material from
the external material during the whole evolution. The main control parameters
are the Mach number of the jet flow, the density contrast between jet and 
external medium and the actual jet density (for a more detailed discussion 
see Paper I). The results that we present are for a single value
of the Mach number $M = 10$ and three different values of the density ratio,
i.e. for an underdense jet ($\nu = 10$), for an equal density jet ($\nu = 1$)
and for an overdense jet ($\nu = 0.1$).  The conservation equations
for mass, momentum and energy are integrated using a numerical code based on
a PPM scheme, on a grid of $256\times 256\times 256$ grid points. The energy equation 
includes also non equilibrium  radiative losses, whose detailed form is 
discussed  in Paper I. 

\section{Results}
In order to determine whether turbulent entrainment that develops as a 
consequence of the evolution of Kelvin-\-Helm\-holtz instabilities   
is a viable mechanism for the acceleration of molecular outflows,
 we will now  compare the properties of the entrained 
material as they result from
 our simulations with some of the typical properties of molecular outflows,
which have been summarized in Sect. 2.
We must notice however that given the nature of our approach, i.e. the temporal
analysis, we cannot address those properties that refer to the distribution
of different quantities along the flow. 

\subsection{Momentum transfer}

The growth of \KH instabilities induces a transfer of momentum from the
jet to the ambient material, which is then accelerated. The efficiency
of this process is very high: Fig. \ref{fig:momentum} shows the behavior
of jet momentum as a function of time for the three cases of overdense,
underdense and equal density jet and we see that, in less than 20 time
units, the jet transfers almost all of its momentum
to the ambient medium (95\%  in the case of the underdense jet). 
Our unit of time is the sound crossing time over the jet radius that 
can be expressed as
$$
 t_{\rm cr} = 334 \ {a_{16}} 
\left( v_{s6} \right)^{-1} {\rm yrs}
$$
 where $a_{16}$ is the jet
radius $a$ expressed in units of $10^{16}$cm and  $v_{s6}$ is the sound
speed $v_s$ expressed is in units of $10^{6}$cm s$^{-1}$. Therefore, if a jet 
carries enough momentum flux to drive a molecular outflow, this mechanism
ensures that it would be transferred almost completely to the ambient
material.

\bef[htbp]

{\includegraphics[width=\hsize,height=5.5cm,bb=120 300 540 650]{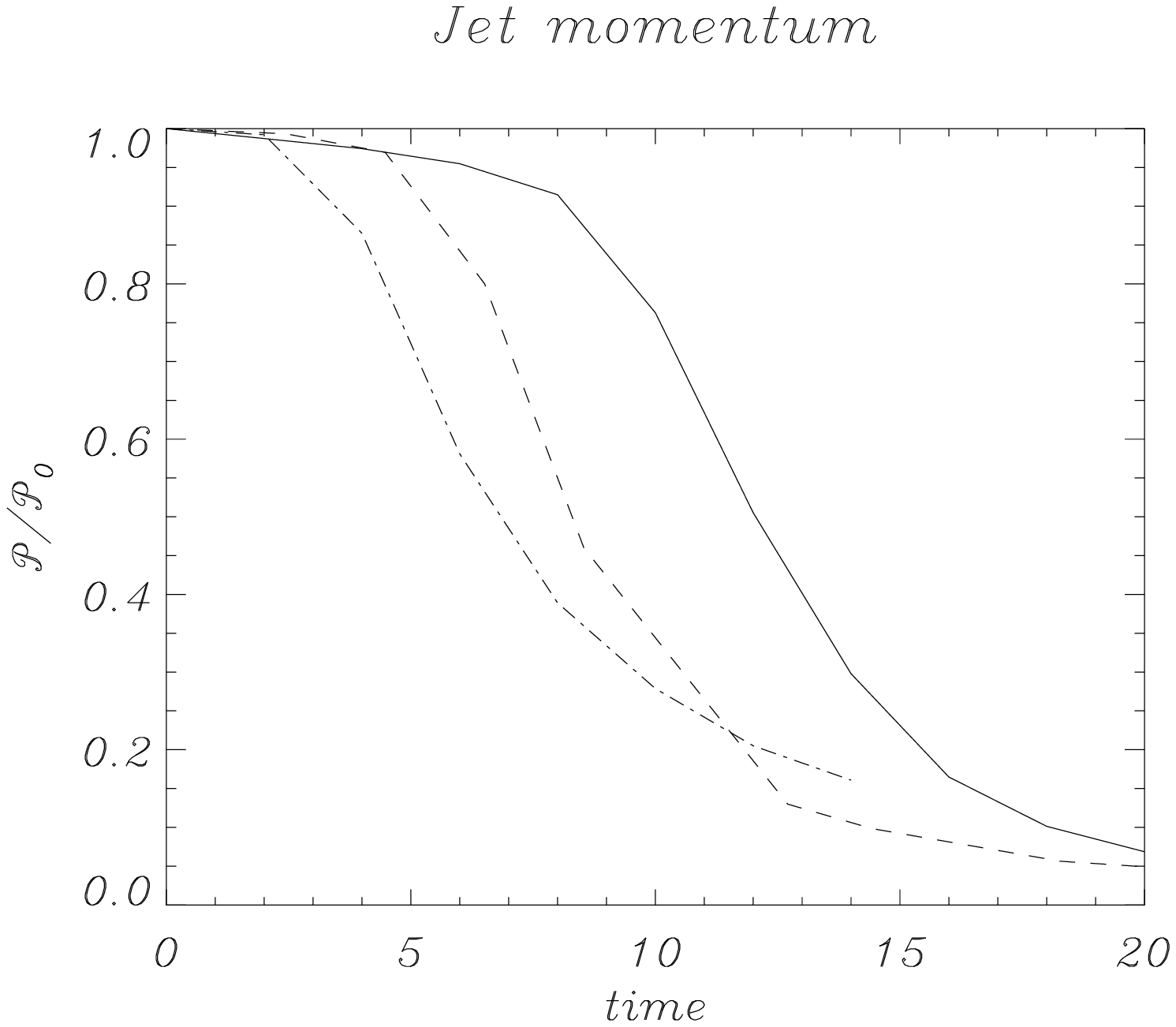}}

\caption{Plot of the jet momentum for heavy (dot-dashed line), 
light (solid line) and equidense jets (dashed line)}
\label{fig:momentum}
\ef  

The next question we can ask is where the momentum is deposited, more
precisely how far from the jet axis the ambient material can be accelerated.
Due to computational limitations, our domain has a limited extension
in the transversal direction, and therefore we cannot follow 
the acceleration of the external material for distances larger than
7 jet radii, nonetheless our data allow us to give some interesting 
estimates. We evaluated the rate of expansion of
the molecular outflow by calculating for each time the
distance $d_\perp$ from the jet axis within which 50\% of the deposited
momentum is contained (Fig. \ref{fig:momflux}).

\bef[htbp]

{\includegraphics[width=\hsize,bb=120 390 540 710]{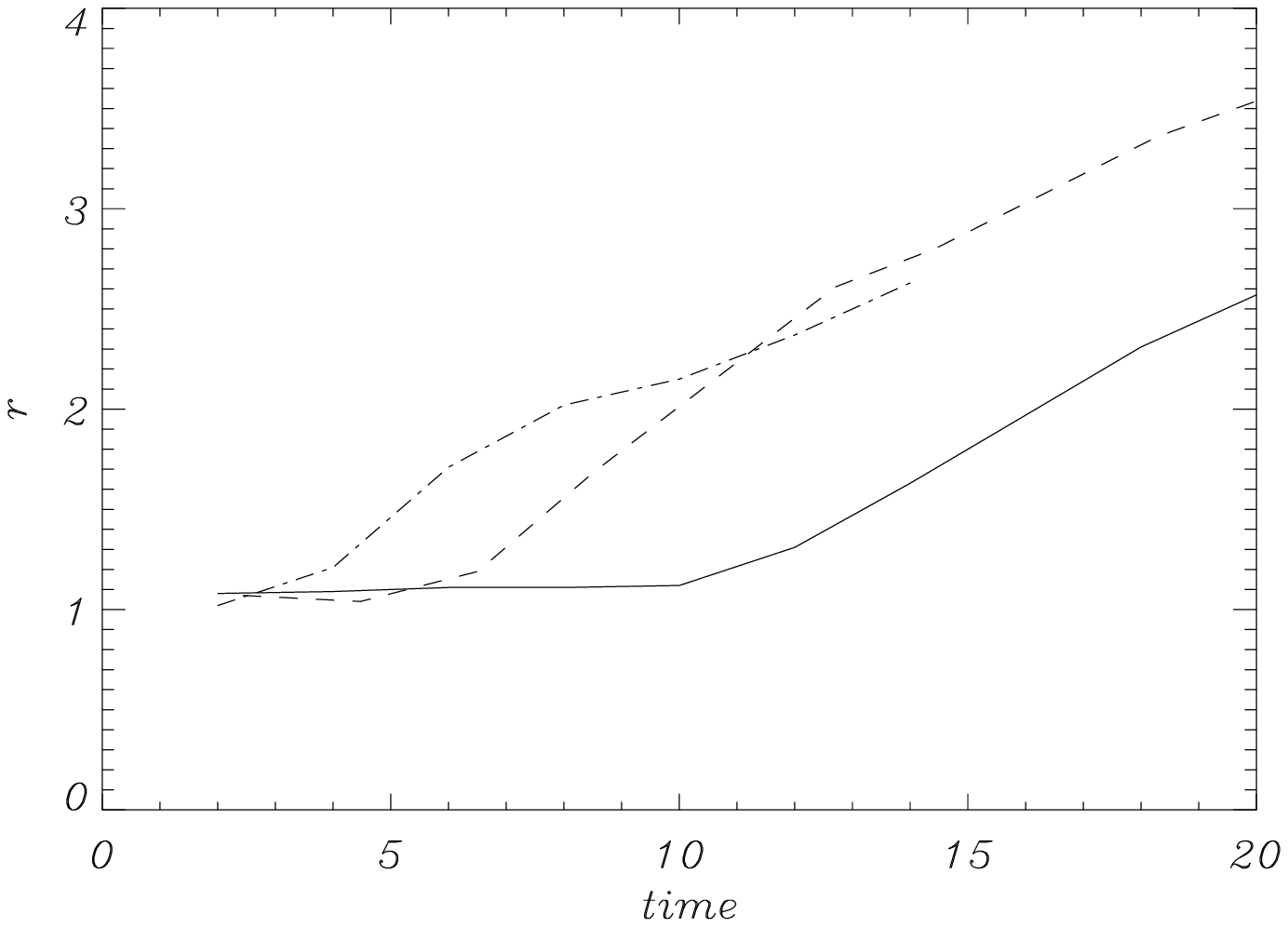}}

\caption{Plot of the distance from the jet radius within which 
50\% of ambient medium momentum is found at each time, for 
heavy (dot-dashed line), 
light (solid line) and equidense jets (dashed line)}
\label{fig:momflux}
\ef  

In the initial times, corresponding to the first stages of 
the instability growth (see Paper I), the ambient material is
accelerated only in the immediate surroundings of the jet; as time
elapses, and the jet starts expanding and mixing with the environment,
the transversal expansion of the molecular outflow grows linearly with 
time, i.e. $d_\perp = v_{\rm exp} t$ where 
the expansion velocity $v_{\rm exp}$ is similar for all cases ($v_{\rm exp} = 0.13 v_s$ for 
heavy jets, $v_{\rm exp} = 0.17 v_s $ for equal-density jets, 
$v_{\rm exp} = 0.15 v_s$ for light jets).
If we extrapolate this linear behavior for longer times
we get, in a timescale of $10^5$yrs, a transverse size of the outflow 
of the order of
$$
d_\perp \sim 4.5 \times 10^{17} {v_{s6} } \  {t_5} \ {\rm cm}
$$
at the lower end of the observed range; ${t_5}$ is the evolutionary time in 
units of $10^5$ yrs.

\subsection{Velocity distributions}

An important constraint that observations provide to theoretical models
is the distribution of the mass of the outflow with velocity.
The variation of outflow emission (and thus of flow mass)
is well represented by a power-law shape $M(v) \propto v^{\gamma}$, 
up to a break velocity,  beyond which the mass decrease is steeper.

We computed the distribution of the entrained mass with velocity
 for all our cases, at all times.
We found that for the light and equal density jet cases, the distribution
of the ambient mass with velocity can indeed be well represented by
a power law, with a break
at high velocity. As an example, we plot, 
in Fig. \ref{fig:mv10},
the distribution of mass with velocity for the light jet case 
at time $t=18$, with the two power law fits superimposed.

\bef[htbp]

{\includegraphics[width=\hsize,bb=100 430 540 750]{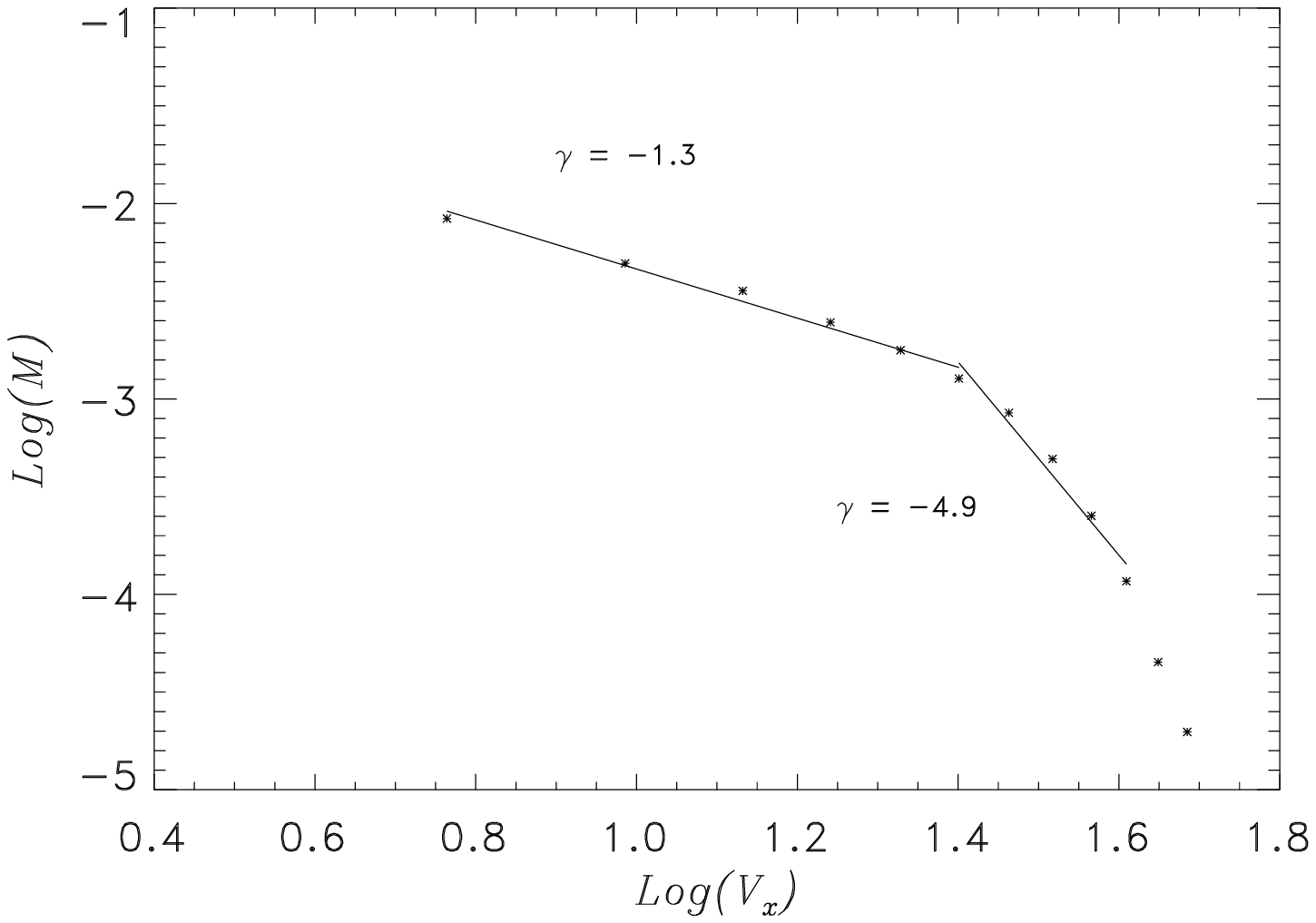}}

\caption{Logarithmic plot of the mass of ambient medium vs 
its velocity for the light jet case at time $t=18$.}
\label{fig:mv10}
\ef 

\bef[htbp]

{\includegraphics[width=\hsize,bb=100 430 540 750]{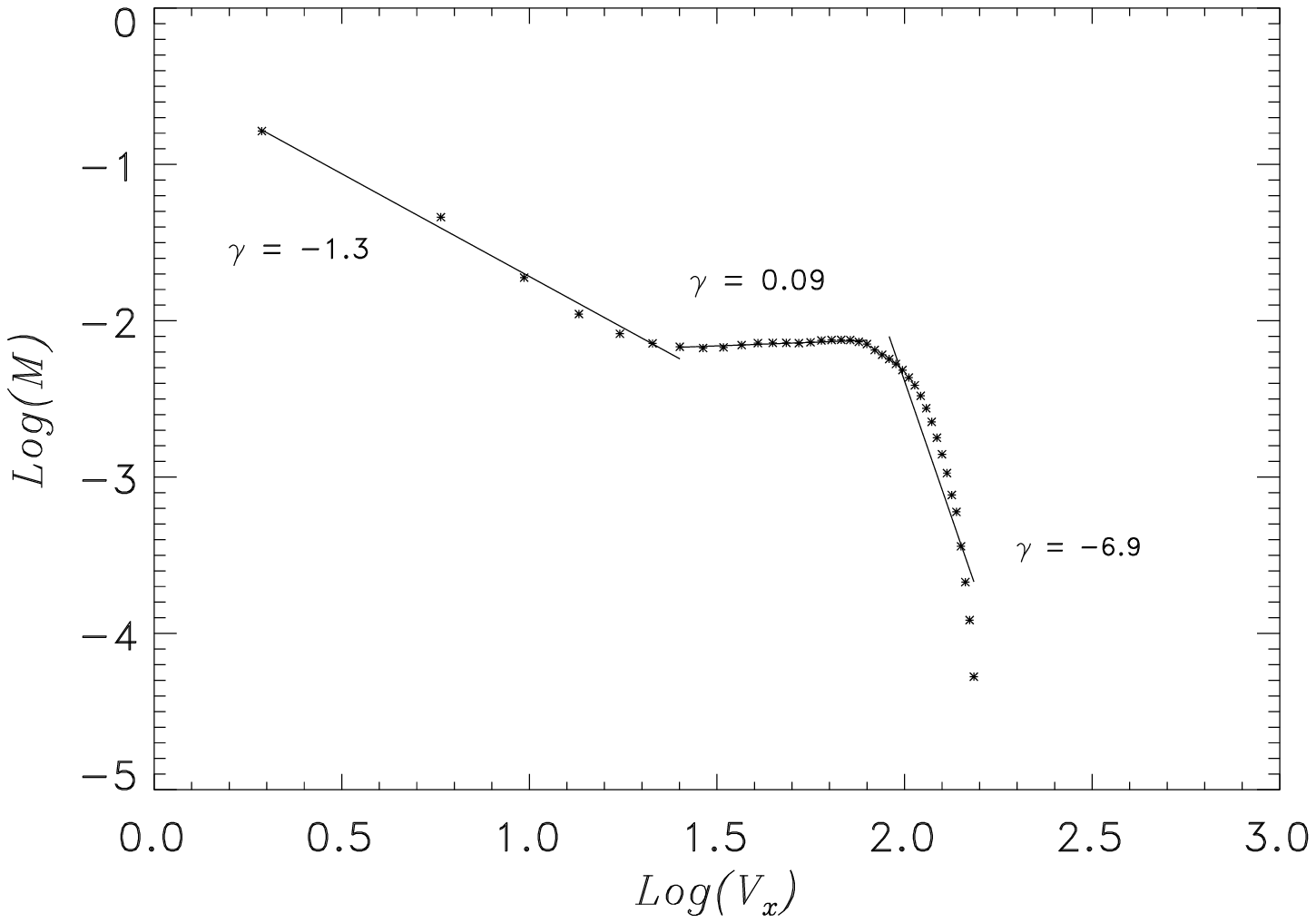}}

\caption{Logarithmic plot of the mass of ambient medium vs 
its velocity for the heavy jet case at time $t=6$.}
\label{fig:mv01}
\ef

In the heavy jet case, instead, the distribution of mass with velocity 
can be represented by three power-law distributions: the mass falls 
steeply with velocity for low and high velocities, while it is almost 
constant or increases at intermediate velocities (Fig. \ref{fig:mv01}).
The intermediate flat part of the distribution tends to become narrower 
with time, however is still present at the end of our simulation. 
It seems, therefore, that the observed distribution cannot be reproduced by
the overdense case, at least in the first part of its evolution.

For this reason we concentrated on the analysis of the other cases and
we examined for them in more detail the behavior  of the slope
of the low velocity region and of the position of the break as a 
function of time, that we plot in Fig. \ref{fig:gammav}.
Two general trends can be recognized: (a) the spectral index of the 
low-velocity region tends to decrease with time, i.e. the low-velocity 
spectrum steepens; 
(b) the break velocity decreases as time elapses.

\bef[htbp]

{\includegraphics[width=\hsize,bb=110 280 540 750]{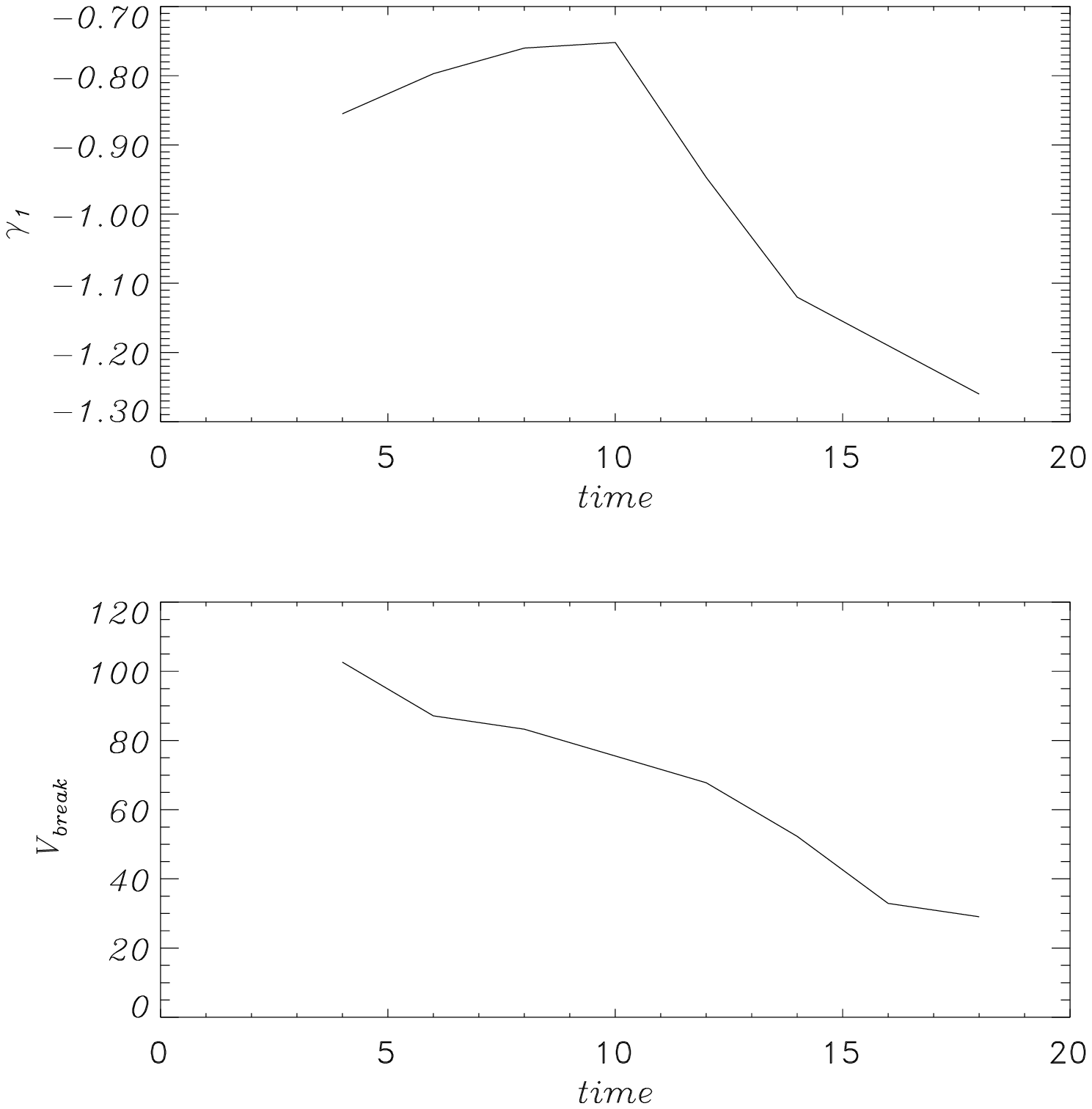}}

\caption{Trend with time of the index $\gamma_1$ (panel a))
and of the break velocity $v_b$ (panel b)) in the power-law 
distribution of ambient mass with velocity $m(v) \propto v^{\gamma}$;
the plots refer to the light jet case.}
\label{fig:gammav}
\ef 

We can interpret the shape and time evolution of the mass distribution 
in terms of
an average velocity profile (Stahler 1994), in fact:

\begin{equation}
{ {dM} \over {dv} } = { {dM} \over {dr} }  
\left( { {dv} \over {dr} } \right)^{-1} =
2 \pi r \rho \left( { {dv} \over {dr} } \right)^{-1} 
\label{eq:stahler}
\end{equation}

where M is the mass per unit length and we have assumed an average homogeneous density distribution.
A power law velocity profile thus corresponds to a power law 
mass distribution: if $v(r) \propto r^{-\alpha}$ we obtain
$dM/dv \propto v^{-1-2/\alpha} $. The actual average velocity profiles 
found in the simulations
can be well represented by power laws in the external
low velocity part, while they flatten in the central 
high velocity region, leading to the steeper part of the 
mass distribution.

In Fig. \ref{fig:velrad} we plot the average velocities for the
dense jet case at a fixed time, and for the
 light jet case at two selected times. 
From this figure, taking into account
the interpretation given above, it is possible
to infer the different behaviour in the mass-velocity distribution 
for heavy and light jet cases. 
For the light jet case (see for example the curve corresponding to time $t=18$
in Fig. \ref{fig:velrad}), the velocity distribution with 
radius is flat at high velocities (0 - $1.5a$) and steeper at intermediate and 
low velocities ($1.5a$ - $3a$).
In the heavy jet case, instead, we have a flat distribution at high velocities
(0 - $0.5a$),
followed by a steeper distribution at intermediate velocities
($0.5a$ - $2a$), while at
low velocities we have a flattening again, reflecting the three components
behavior of the mass distribution.

\bef[htbp]

{\includegraphics[width=\hsize,bb=110 490 530 730]{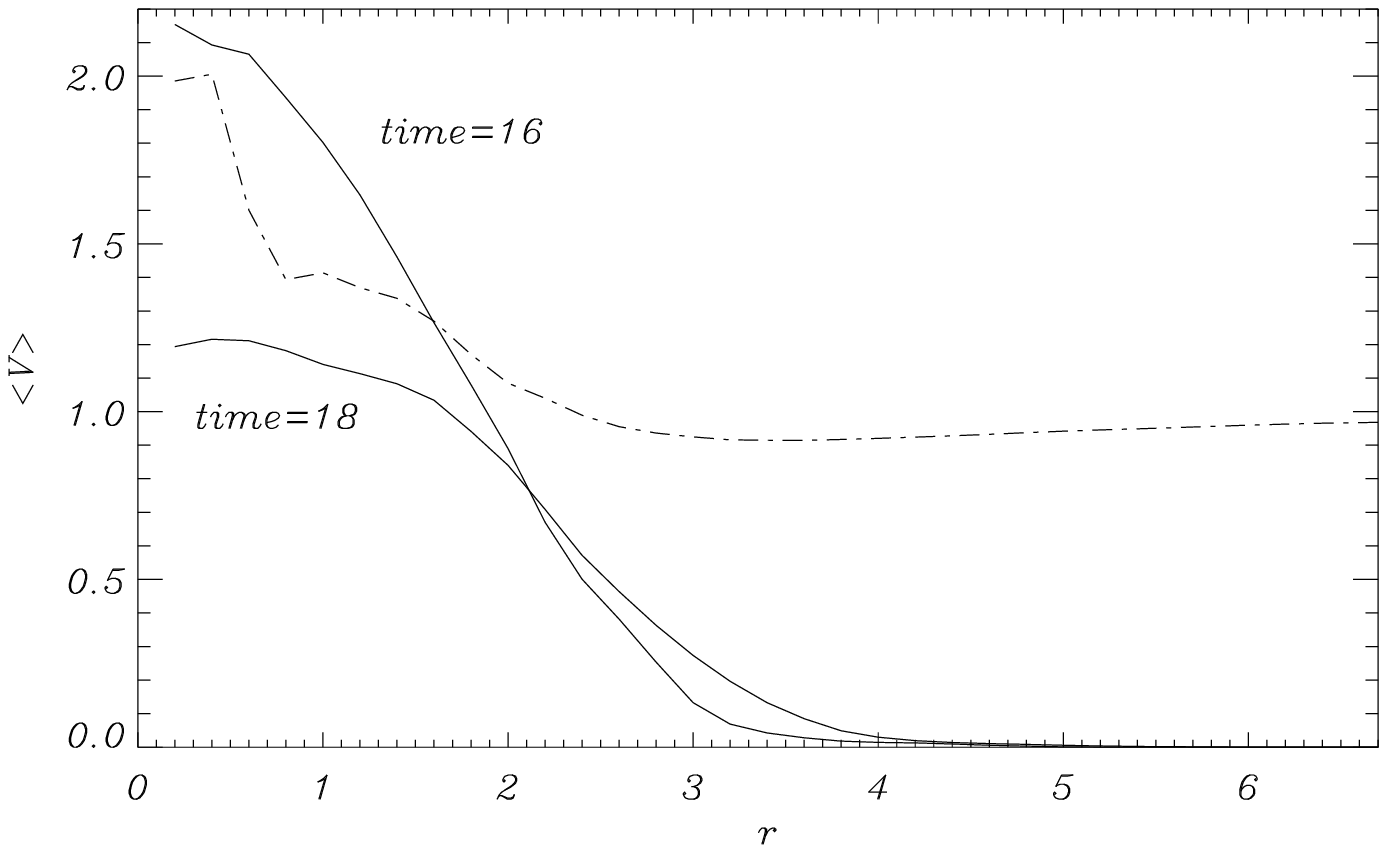}}

\caption{Average distribution of the velocity of the ambient medium
versus the distance from the jet axis. The solid lines are for the
light jet case, the dot-dashed line is for the heavy jet case.
The velocity distribution for the heavy jet case has been multiplied 
by 10 in order to facilitate the comparison with the other case.}
\label{fig:velrad}
\ef 

\bef[htbp]

{\includegraphics[width=\hsize,bb=40 250 310 690]{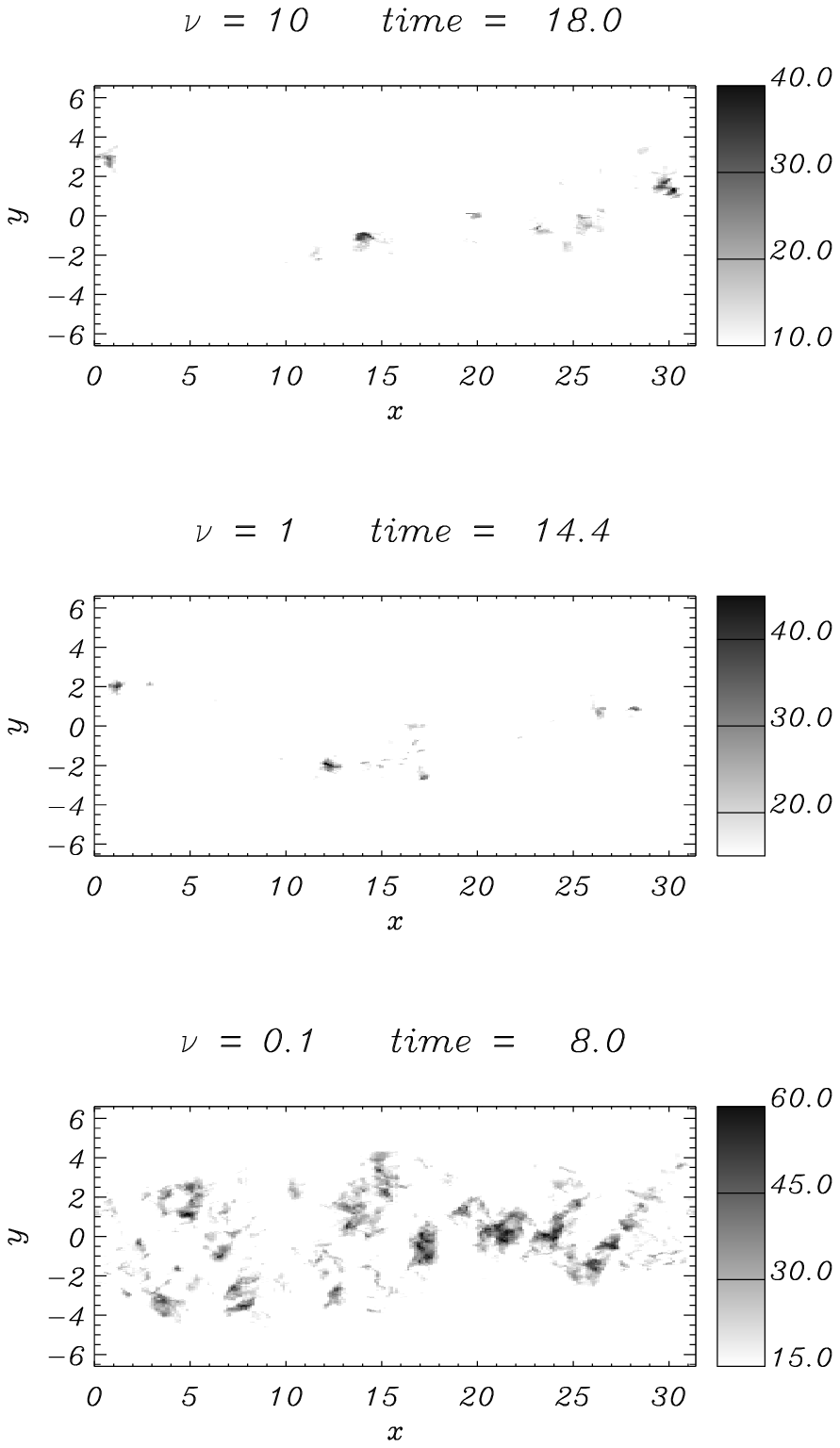}}

\caption{High velocity structures in the ambient medium, integrated
over the $z$ direction.}

\label{fig:bullets}
\ef

Considering the time evolution of the mass distribution, we can deduce the 
temporal variation of the break velocity and of the spectral index
(Fig. \ref{fig:gammav}) from the temporal variation of the distribution
of the average velocity with radius. In fact, comparing the two curves in 
Fig. \ref{fig:velrad} referring to the light jet case at times $t=16$
and $t=18$ (solid lines), we find that the maximum velocity 
value decreases with time, as well as the ``break" value where the 
distribution of velocity  with radius changes its slope
($\langle V \rangle \approx 2, \ t=16; \ \langle V \rangle \approx 1, \ t=18$). 
This explains the  decrease in the break velocity with time, which 
is a common trend in all the studied cases (Fig. \ref{fig:gammav}, panel b)).
Moreover, comparing the slope of the average velocity profile at time
$t=16$ with the one at time
$t=18$ in the low velocity range, between $2a$ and $\sim 3a$  
(Fig. \ref{fig:velrad}), 
we note that the velocity distribution at time $t=18$ is
flatter since more material has been accelerated at larger
distances from the outflow axis.
This implies a steepening
of the mass distribution at low velocities, according to Eq. \ref{eq:stahler},
and this is indeed what we 
see in the light and equal density jet cases (Fig. \ref{fig:gammav}, panel a)).

Although the average velocity has the smooth profile shown in Fig. 
\ref{fig:velrad}, the actual distribution 
can however show high velocity peaks, which may be correlated with
the so called ``EHV bullets".  We have marked the material moving 
at $V>40$ km s$^{-1}$ and integrated over the $z$ direction. 
The resulting images are displayed in Fig. \ref{fig:bullets}.
The figure shows  that turbulent entrainment is 
able to accelerate ``bullets" of material at high velocities and these 
are found within a few radii from the jet axis.

\subsection{Collimation}

Observational data on molecular outflows show that the flow collimation
increases with flow velocity.
Moreover, the blue-shifted and red-shifted lobes are generally well separated,
especially at high velocities.

\bef[htbp]

{\includegraphics[width=\hsize,bb=80 90 570 740]{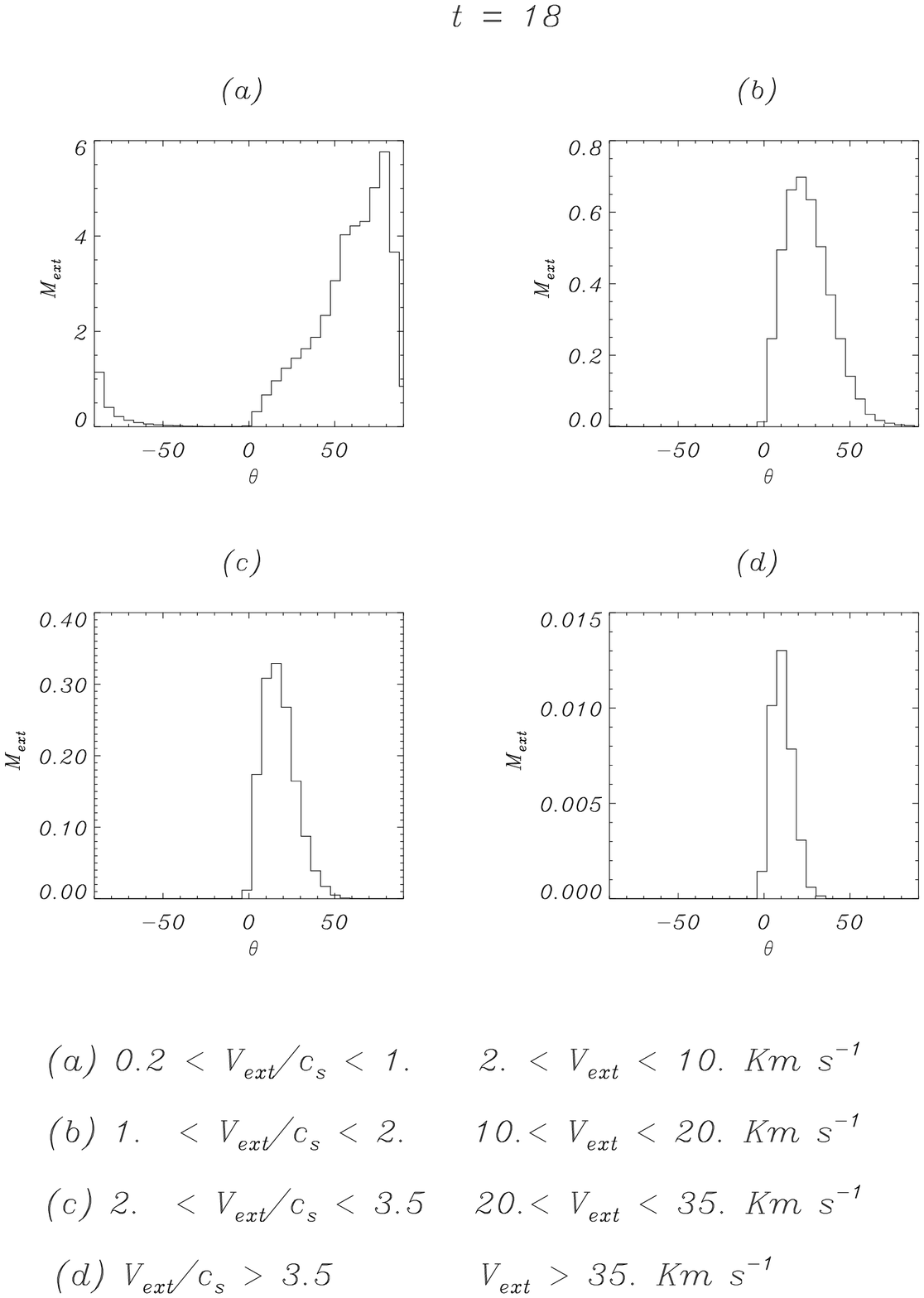}}

\caption{Distribution, in four different velocity intervals, of the ambient
medium mass with respect to the angle $\theta$ between the flow velocity
and the initial jet axis. These plots concern the light jet case,
at time $t=18$}
\label{fig:mtheta}
\ef

To verify whether these features are reproduced in our calculations, 
we considered at each grid point the angle $\theta$ between the 
outflow velocity and the initial jet axis. Smaller $\theta$ imply more 
collimated material.
We divided the observed velocity range in four intervals, and 
for each of them we calculated the distribution of the 
ambient material mass with respect to $\theta$. 
One example of the distributions  that we obtained 
is shown in Fig. \ref{fig:mtheta}, for the light jet case, at time 
$t=18$.

At low velocity the flow angle is distributed over a wide interval, and
thus the flow is poorly collimated. The material moving at higher velocities 
moves at smaller angles respect to the initial flow axis, and is thus more 
collimated, in agreement with observations.
This behavior is observed in all the studied cases, although a jet moving
in a denser environment accelerates a more collimated outflow, while
dense jets accelerate less collimated outflows.
This trend can be inferred from Table \ref{tab:mass90} where we report
the value of the angle $\theta$ within which is contained 90\% 
of the ambient mass whose velocity lies in one of the 
four intervals defined in Fig. \ref{fig:mtheta}, at three significative
times for the three cases. The variation with time is in any case quite small.

\begin{table}[htb]

\begin{center}
\begin{tabular}{|c|c|c|c|}  \hline
 & $\nu = 10$ & $\nu = 1$  & $\nu = 0.1$ \\
 & $time=18$ &  $time=18$ & $time=12$ \\
\hline

$2<V<10$ km s$^{-1}$  &  88$^\circ$   & 101$^\circ$    &  124$^\circ$     \\
\hline

$10<V<20$ km s$^{-1}$ &  48$^\circ$   &  59$^\circ$   &   96$^\circ$   \\
\hline

$20<V<35$ km s$^{-1}$ &  36$^\circ$   &  36$^\circ$   &  48$^\circ$    \\
\hline

$V>35$ km s$^{-1}$ &  24$^\circ$   &     &   30$^\circ$    \\
\hline

\end{tabular} \end{center}

\caption{Angle containing 90\% of the ambient mass moving with 
velocity in the selected ranges, for the three cases, at three
significative times.} 

\label{tab:mass90}
\end{table}

\bef[htbp]

{\includegraphics[width=\hsize,bb=90 140 550 710]{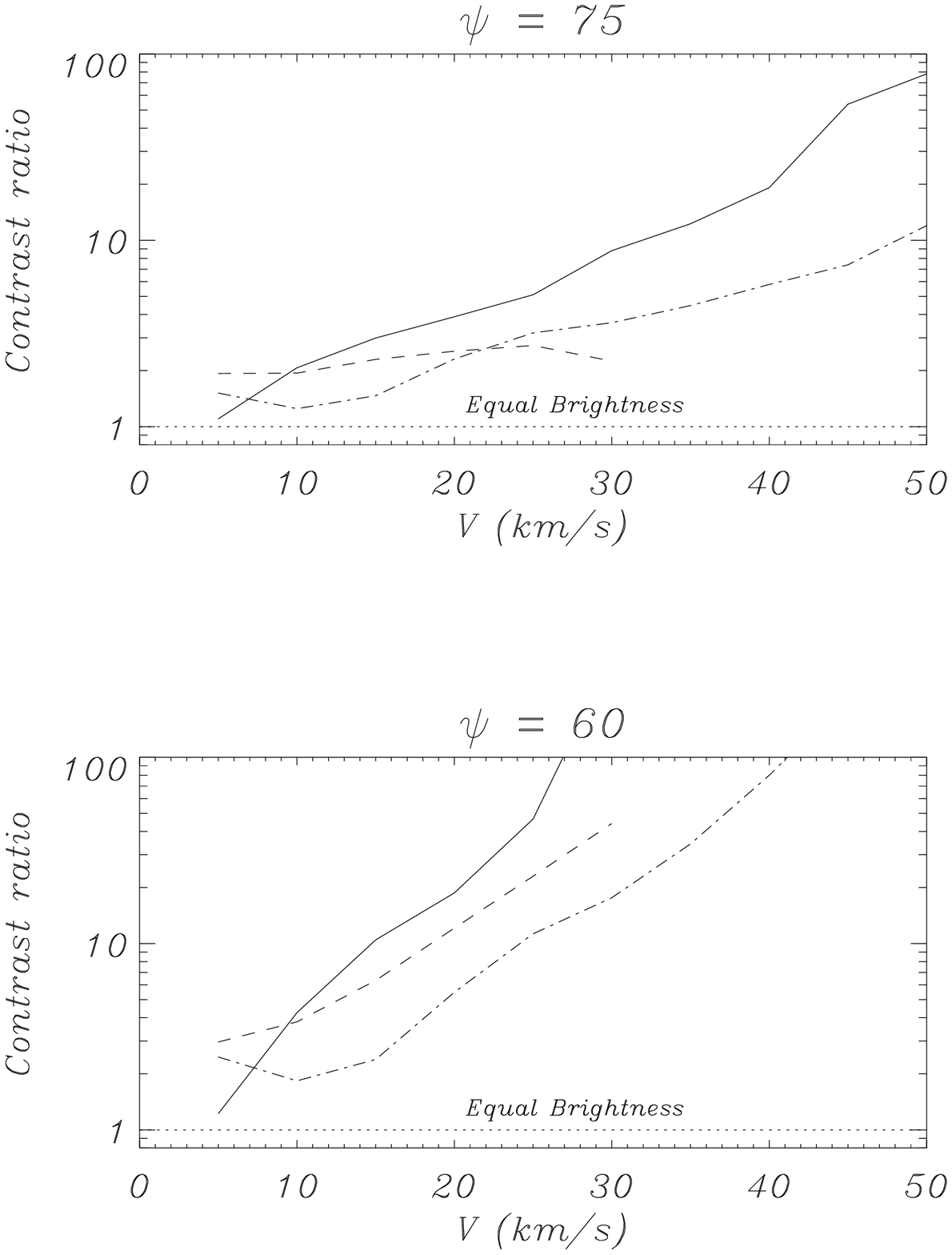}}

\caption{Ratio between the mass of blue-shifted and red-shifted material,
in the blue-shifted lobe, for outflows moving at an angle of $75^\circ$
(upper panel) and $60^\circ$ (lower panel) respect to the line of view,
versus the flow velocity. The solid, dashed and dot-dashed lines refer,
respectively,  to the light, equal-density and heavy jet cases.}

\label{fig:blured}
\ef 

We evaluate now how these results compare with
the observations  regarding the overlapping of blue-shifted
and red-shifted emission in molecular outflows.

This obviously depends on the angle $\psi$ between the line of sight 
and the flow axis (coincident with our $x$ axis) 
and we choose for our calculation two values for
this angle, i.e.  $\psi=75^\circ$ and
$\psi=60^\circ$. For these two angles we compute the ratio of the 
blue-shifted (positive velocity along the line of sight)
to the red-shifted (negative velocity along the line of sight) mass.
In Fig. \ref{fig:blured} we plot this ratio as a function of velocity.
We notice that the ratio increases as this angle diminishes; 
when the flow is more directed towards the observer, in fact,
only the material moving with large angles respect to the flow axis
(i.e. the loosely collimated material) will appear red-shifted.

From our results we notice that the contrast ratio between blue-shifted
and red-shifted material increases with velocity (again, the faster 
material is the more collimated one), in agreement with the data
of, eg. Lada \& Fich (1996). The contrast ratio, at all velocities,
is higher for outflows driven by light jets, compared to outflows 
driven by equal density and heavy jets.
In Fig. \ref{fig:blured} we plot the trends for three selected times,
and this is possible since we did not notice significant differences in the 
values and trends of the contrast ratios at different evolutionary times.

\subsection{Acceleration and dissociation}

One problem of the prompt entrainment
mechanism for the acceleration of molecular outflows, is that as ambient
medium is accelerated by the passage of a bow shock, it is also 
dissociated (Downes \& Ray 1999), and thus, although the bow shock
is very efficient in transferring momentum, most of it
goes to atomic material and  the effective fraction of momentum
transfered by the jet to molecular material is low.

\bef[htbp]

{\includegraphics[width=\hsize,bb=80 40 570 740]{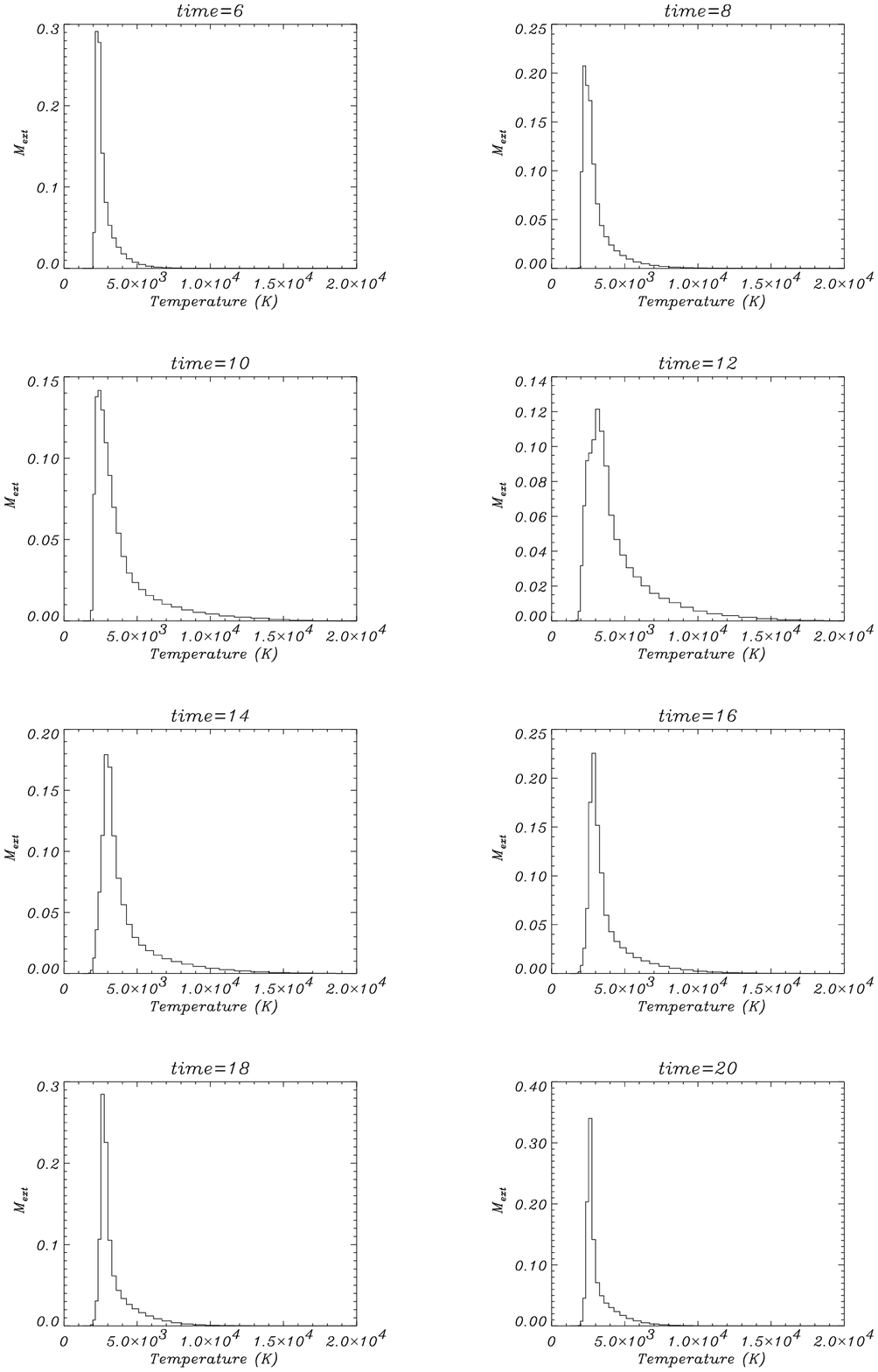}}

\caption{Distribution of the ambient accelerated mass as a function of 
its temperature, for different times, and for the light jet case.}
\label{fig:tmass}
\ef 

In our case we cannot discuss molecular dissociation, since we 
did not include in our computations 
all the physics related to molecular formation and dissociation, and to
molecular radiative losses.
However it is possible to have a feeling of what the situation would be, 
by considering the temperature of the accelerated material.
We limit the discussion to the light jet case: since we start from
a configuration where the jet is in pressure equilibrium with the
ambient medium, in the $\nu=10$ case the initial external temperature
is $1/10$ of the initial jet temperature (i.e. $T_{ext}=1000$ K),
while in the other cases the ambient medium is hotter from the beginning
of the calculations and cannot be molecular.

In Fig. \ref{fig:tmass} we plot the distribution of the accelerated ambient 
material (i.e. external material with velocity $V>2$ km s$^{-1}$)
with temperature;
we note that at intermediate times 
(eg. $t=10,12,14$) a fraction of
the ambient material is heated, and this is consistent with the 
fact that at this stage strong shocks are driven by the jet into the 
external medium. As shocks become weaker and disappear, 
the temperature of the accelerated ambient medium decreases.
At all times, however, the largest fraction of accelerated material
maintains a temperature not too different from the initial value.
We can therefore expect that the ambient medium accelerated through 
turbulent entrainment at the jet surface remains mostly molecular.

\section{Summary and Conclusions}

We have shown that turbulent mixing in Kelvin-Helmholtz unstable jets is a very
efficient mechanism of momentum transfer from the jet to the ambient medium:
at the final stages of the evolution, up to 95\% of the jet momentum has 
been transfered to entrained ambient material; since the momentum carried
by parsec-scale YSO jets is found to be comparable to the momentum of typical 
molecular outflows, we infer that this mechanism alone can be invoked to 
explain the observed outflows. Moreover, opposite to what happens in bow-shock
accelerated material, the temperature of the accelerated ambient medium 
remains low, and thus momentum is actually acquired by molecular material.

The transverse size of the outflows that are generated 
in this way is located at the lower end of the observed range, and our
length to width ratio is  too high with respect to the average value 
for molecular outflows but it is  similar to that of highly collimated molecular
outflows such as, for example NGC 2264G, Mon R2 and RNO43.

An interesting feature of our model is that it reproduces the observed 
distribution of mass with velocity: it has the form of a power law
$M(v)\propto v^{\gamma}$, with a break at high velocities. The shape of
the distribution and the values of the spectral index and of the break 
velocity are particularly well reproduced by the case in which a light 
jet is moving into a denser environment. 
The most powerful observed molecular outflows are located in the inner 
regions of molecular clouds, where the ambient medium is often so dense that 
the optical jets are completely obscured. It is thus plausible that in these 
cases the atomic jets are much lighter than the environment, as implied by 
our model. Conversely, bright optical jets coming out of the parent clouds,
that would be probably denser than the surrounding ambient material, are 
accompanied by very weak molecular outflows, if any 
(HH 34 and HH 1-2, Chernin \& Masson 1995). 
One exception is the bright optical jet HH 111, which is associated
to a powerful molecular outflow (Cernicharo \& Reipurth 1996). However,
also in this case, the blueshifted jet breaking through the surface 
of the cloud is accompanied by a weaker molecular outflow with respect to the 
redshifted lobe surrounding the optically obscured counter-jet  
(Reipurth \& Olberg 1991).

As far as the distribution of velocity with distance 
from the source is concerned, 
we could not reproduce the observed trend: we found that the average velocity 
of the outflow decreases with time (and thus with distance from the source, 
although only qualitative considerations can be performed on quantities requiring
a translation from a temporal to a spatial approach), and this behavior
is opposite to the observed ``Hubble Law". 
To explain the increase in velocity with distance from the source, 
many possibilities have been invoked: in the frame of the steady state model, 
Stahler (1994) inferred from analytical calculations that the Hubble law 
was a consequence of material with progessively higher velocities coming into
view at greater distances from the star.
The Hubble law could possibly be  due to a decreasing density of the ambient medium
(Padman et al. 1997), although, more probably,  
high velocities at the head of the flow are actually due to the presence of 
the bow shock that accelerates ambient material through prompt entrainment.

The collimation and bipolarity properties of the outflows generated with 
Kelvin Helmholtz 
induced turbulent entrainment fit very well the observations: 
the flow collimation increases with velocity, and there is a clear separation
of  the blue-shifted and red-shifted material, which increases as well with 
velocity. An example of a molecular outflow for which this behavior has been studied
in detail is NGC 2264G, Lada \& Fich (1996).

Since these preliminary results are encouraging,
future work will analyze the combined effects of the two mechanisms,
the turbulent entrainment at the jet surface and the prompt entrainment at 
the bow shock, studying the propagation of the jet's head in space, 
and  contemporarily the spatial growth of the Kelvin-Helmholtz unstable modes
in the jet body.

\bigskip

{\it Acknowledgments:} The calculations have been per\-for\-med on the Cray T3E 
at CINECA in Bologna, Italy, thanks to the support of CNAA. 
This work has been supported in part by the DOE ASCI/Alliances grant at the
University of Chicago. M.M. acknowledges the CNAA 3/98 grant. 

\section{References}

\noindent
Bachiller, R., 1996, Ann.Rev.Astron.Astrophys 34, 111

\noindent
Bally, J.,  Devine, D., 1994, ApJ Lett. 428, 65

\noindent
Bodo, G., Massaglia, S., Ferrari, A., Trussoni, E., 1994, A\&A 283, 655 

\noindent
Bodo, G.,Massaglia, S., Rossi, P.,  Rosner, R., Malagoli, A., 
Ferrari, A., 1995, A\&A 303, 281

\noindent
Bodo, G., Rossi, P., Massaglia, S., Ferrari, A., 
Malagoli, A., Rosner, R.,  1998, A\&A 333, 1117

\noindent
Cabrit, S., Raga, A., Gueth, F., 1997, in ``Herbig-Haro Outflows
and the Birth of Low Mass Stars", B. Reipurth \& C. Bertout eds.,
IAU Symposium No. 182, Kluwer Academic Publishers, 163

\noindent
Cernicharo, J., Reipurth, B., 1996, ApJ 460, L57

\noindent
Chernin, L.M., Masson, C.R., 1995A, ApJ 455, 182

\noindent
Chernin, L.M., Masson, C.R., 1995B, ApJ 443, 181

\noindent
Davis, C.J., Eisloeffel, J., Ray, T.P., Jenness, T.,  1997, 
A\&A 324, 1013

\noindent
Downes, T.P.,  Ray, T.P.,  1999, A\&A 345, 977

\noindent
Fukui, Y., Iwata, T., Mizuno, A., Bally, J., Lane, A.P., 1993, in
``Protostars and Planets III", E.H. Levy \& J.I. Lunine eds., 
University of Arizona Press, Tucson, 603

\noindent
Lada, C.J.,  1985, Ann.Rev.Astron.Astrophys 23, 267

\noindent
Lada, C.J., Fich, M.,  1996, ApJ 459, 638

\noindent
Lizano, S., Heiles, C., Rodriguez, L.F., Koo, B., Shu, F.H., 
Hasegawa, T., Hayashi, S., Mirabel, I.F., 1988, ApJ 328, 763

\noindent
Margulis, M., Lada, C.J., Hasegawa, T., Hayashi, S., Ha\-ya\-shi, M.,
Kaifu, N., Gatley, I., Greene, T.P., Young, E.T.,  1990, ApJ 352, 615

\noindent
Masson, C.R., Chernin, L.M.,  1992, ApJ Lett 387, 47

\noindent
Meyers-Rice, B.A.,  Lada, C.J.,  1991, ApJ 368, 445

\noindent
Micono, M., Massaglia, S.,  Bodo, G., Rossi, P., Ferrari, A., 1998,
{\it A\&A} {\bf 333}, 989

\noindent
Micono, M., Bodo, G., Massaglia, S., Rossi, P., Ferrari, A., Rosner, R., 2000,
A\&A, in press

\noindent
Moriatry-Schieven, G.H., Snell, R.L.,  1988, ApJ 332, 364

\noindent
Moriatry-Schieven, G.H.,  Hughes, V.A., Snell, R.L., 1989, ApJ 347, 358

\noindent
Padman, R., Bence, S., Richer, J.,  1997, in ``Herbig-Haro Outflows
and the Birth of Low Mass Stars", B. Reipurth \& C. Bertout eds.,
IAU Symposium No. 182, Kluwer Academic Publishers, 123

\noindent
Parker, N.D.,  Padman, R., Scott, P.F., 1991, MNRAS 252, 442

\noindent
Raga, A., Binette, L., Cant\'o, J., 1990, ApJ 360, 612

\noindent
Reipurth, B., Olberg, M., 1991, A\&A 246, 535

\noindent
Richer, J.S., Hills, R.E., Padman, R., 1992, MNRAS 254, 525

\noindent
Rodriguez-Franco, A., Martin-Pintado, J., Wilson, T.L., 1999, A\&A 351, 1103

\noindent
Rossi P., Bodo, G., Massaglia, S., Ferrari, A., 1997, A\&A
321, 672  

\noindent
Stahler, S.W., 1994, ApJ 422, 616

\noindent
Taylor, S.D., Raga, A.C., 1995, A\&A 296, 823

\noindent
Uchida, Y., Kaifu, N., Shibata, K., Hayashi, S., Hasegawa, T., Hamatake, H., 1987, 
Publ. of the Astron. Soc. of Jap., 39, 907

\end{document}